%% file: arxiv.tex
\documentclass[numbers,sort&compress]{article}
\usepackage{graphicx}
\usepackage{listings}
\usepackage{url}
\usepackage{lineno}
\usepackage{subcaption}
\usepackage{rotating}
\usepackage[normalem]{ulem}
\usepackage{arxiv}
\usepackage{amssymb,amsmath,amsthm}
\usepackage[numbers,sort&compress]{natbib}
\usepackage{xcolor}
\usepackage{hyperref}
\usepackage{tabularx}
\usepackage{mhchem}
\usepackage[toc,page]{appendix}
\graphicspath{{Artworks/}{imgs/}}
\usepackage{amsfonts}
\usepackage{nicefrac}
\usepackage{microtype}
\usepackage{cleveref}
\usepackage{doi}
\usepackage{tikz}
\usetikzlibrary{positioning,arrows.meta,fit,backgrounds,calc,shapes.geometric}
\definecolor{RuhiTeal}{HTML}{004D40}
\definecolor{RuhiCoral}{HTML}{FF655D}
\definecolor{RuhiYellow}{HTML}{F1DB4B}
\input{authors_arxiv.tex}
\date{\today}
\title{Reproducible Orchestration of Best Practices for Reaction Path Optimization with the Nudged Elastic Band}
\hypersetup{
 pdfauthor={},
 pdftitle={Reproducible Orchestration of Best Practices for Reaction Path Optimization with the Nudged Elastic Band},
 pdfkeywords={},
 pdfsubject={},
 pdfcreator={Emacs 30.2 (Org mode 9.7.39 + HaoZeke)}, 
 pdflang={English}}
\begin{document}

\maketitle
\begin{abstract}
The nudged elastic band (NEB) method is the standard approach for finding minimum energy paths and transition states on potential energy surfaces. Practical NEB calculations require several pre-processing steps: endpoint minimization, structural alignment, and initial path generation. These steps are typically handled by ad-hoc scripts or manual intervention, introducing errors and hindering reproducibility. We present a fully automated, open-source Snakemake workflow for small gas phase molecules that couples modern machine learning potentials (PET-MAD) to the eOn saddle point search software. Each step of the calculation lifecycle is encoded as an explicit dependency graph, from model retrieval and endpoint preparation through path initialization and band optimization. The workflow resolves all software dependencies from conda-forge, ensuring identical execution across platforms. Validation on the HCN to HNC isomerization demonstrates that the automated pipeline recovers the known single-barrier energy profile and product energy without manual intervention.
\end{abstract}
\keywords{Reaction Path Optimization; Snakemake; Nudged Elastic Band;
Reproducibility; Workflow Automation; High Performance Computing}
\section{Introduction}
\label{sec:org2599d16}

The Nudged Elastic Band (NEB) method \cite{jonssonNudgedElasticBand1998} finds the
Minimum Energy Path (MEP) or the path of highest statistical weight in
configuration space by optimizing a chain of \(N\) images connecting reactant and
product endpoints. Each image experiences a projected potential force
perpendicular to the path and a spring force parallel to it; the resulting
``nudged'' force prevents corner-cutting while maintaining image spacing. The
climbing image modification \cite{henkelmanClimbingImageNudged2000} drives the
highest-energy image to the first-order saddle point, and energy-weighted
springs \cite{asgeirssonNudgedElasticBand2021} concentrate images near the barrier
region. Modern implementations combine band optimization with single-ended
saddle search refinement for higher accuracy at the transition state
\cite{goswamiEfficientExplorationChemical2025,goswamiEnhancedClimbingImage2026}
or even switch to locally approximated surfaces
\cite{goswamiEfficientImplementationGaussian2025,goswamiAdaptivePruningIncreased2025,petersonAccelerationSaddlepointSearches2016}.

In practice, NEB calculations require careful preprocessing: endpoint
structures must be minimized, atom orderings aligned, and an initial path
generated that avoids atomic collisions. Mismatched atom labels or poor initial
guesses cause outright failure or convergence to irrelevant saddle points.
These steps are seldom integrated into NEB codes, leaving users to assemble
ad-hoc scripts that are difficult to reproduce.

We present an open-source Snakemake \cite{molderSustainableDataAnalysis2021}
workflow that automates this entire lifecycle, with all dependencies resolved
from conda-forge via \texttt{pixi}. The workflow encodes each step as an explicit node in
a dependency graph: endpoints are minimized and aligned, an initial path is
generated via pairwise-potential interpolation, and only then does the NEB
optimization begin. Users supply endpoint structures and a configuration file;
the workflow handles the rest.
\section{The NEB Orchestrator Workflow}
\label{sec:orgbf676ea}

\subsection{Workflow as a Directed Acyclic Graph}
\label{sec:orgb2da106}

\begin{figure}[t]
  \centering
  \includegraphics[width=\textwidth]{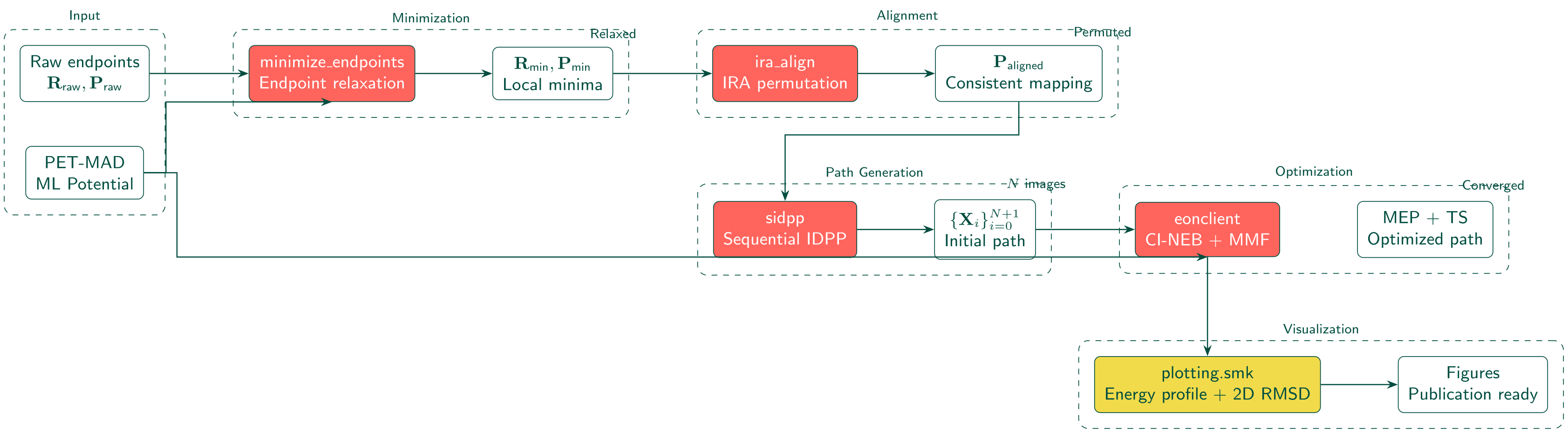}
  \caption{Workflow pipeline showing the Snakemake dependency graph. Raw
  endpoints undergo minimization and IRA alignment before SIDPP path generation.
  The hybrid CI-NEB-MMF optimization then refines the path to the MEP and
  transition state.}
  \label{fig:workflow}
\end{figure}

The workflow is encoded as a directed acyclic graph (DAG) \(G = (V, E)\) where
vertices represent rules (\texttt{get\_model}, \texttt{minimize}, \texttt{align}, \texttt{idpp}, \texttt{neb}, \texttt{visualize}) and
edges represent data dependencies (Figure \ref{fig:workflow}). Three properties
follow directly:

\begin{description}
\item[{Automatic parallelization}] Independent rules execute concurrently, e.g.
\(\text{minimize}(\mathbf{R}) \parallel \text{minimize}(\mathbf{P})\).
\item[{Incremental recomputation}] Modifying the reactant re-runs downstream rules
but skips model retrieval and product minimization.
\item[{Enforced ordering}] \(\text{neb} \leftarrow \text{idpp} \leftarrow
  \text{align} \leftarrow \text{minimize} \leftarrow \text{raw inputs}\).
\end{description}

Snakemake traverses this graph, executing rules only when dependencies are
satisfied and inputs have changed.
\subsection{Automated Dependency Management}
\label{sec:org993bbca}

Beyond code, reproducibility requires pinning the workflow itself, including the
models. The \texttt{get\_model.smk} rule fetches a specific version of the PET-MAD
\cite{mazitovPETMADLightweightUniversal2025} checkpoint from HuggingFace
(\texttt{lab-cosmo/upet}), converts it to a deployable \texttt{.pt} file via \texttt{mtt export}
(metatrain), and marks the output as protected against accidental deletion.
Alternative models are supported through the same metatomic interface
\cite{bigiMetatensorMetatomicFoundational2026}, requiring only a configuration
change. Other models supported by eOn \footnote{\url{https://eondocs.org}} are also a
parameter away.
\subsection{Endpoint Preparation}
\label{sec:orgf14342b}

Consistent atom-to-atom mapping between reactant and product is essential. The
workflow implements a two-stage preparation process.
\subsubsection{Minimization}
\label{sec:orgdd6bb2e}

The \texttt{minimize\_endpoints} rule relaxes each endpoint structure to the nearest local
minimum on the chosen potential energy surface, preventing the NEB band from
stalling on steep gradients at its ends. Configuration parameters:

\begin{description}
\item[{Optimizer}] Limited memory Broyden–Fletcher–Goldfarb–Shanno \cite{nocedalUpdatingQuasiNewtonMatrices1980}
\item[{Force convergence}] 0.0514221 eV/\AA{}$\backslash$ (10\(^{-3}\) Ha/Bohr)
\item[{Maximum iterations}] 2000
\item[{Maximum movement}] 0.1 \AA{}
\end{description}
\subsubsection{Alignment}
\label{sec:org8330ee6}

\begin{figure}[t]
  \centering
  \includegraphics[width=0.8\textwidth]{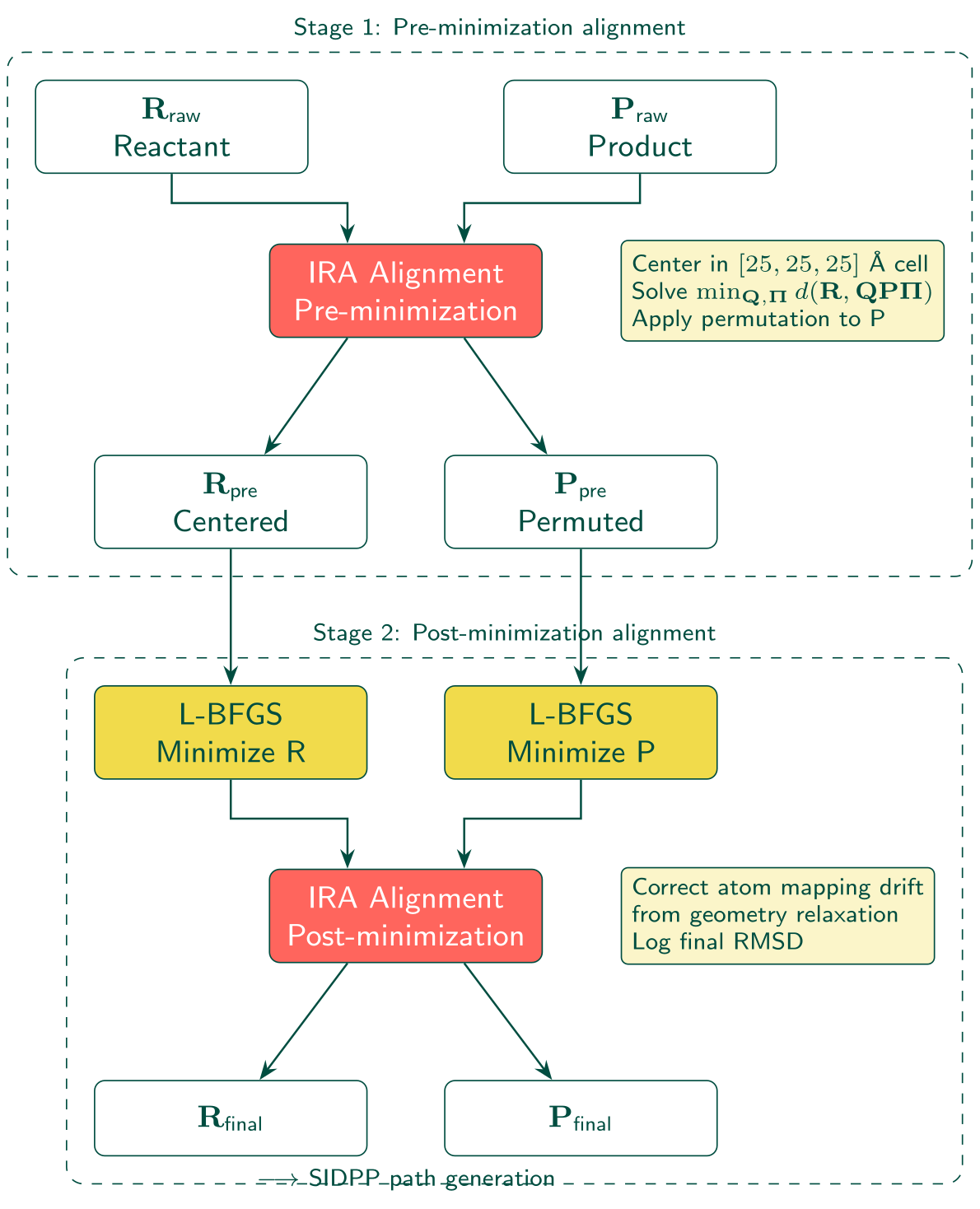}
  \caption{Two-stage IRA alignment process. Raw endpoints are centered and
  aligned before minimization to establish consistent atom ordering. After
  geometry relaxation, alignment is reapplied to correct any atom mapping drift
  introduced during optimization. Both stages solve the joint rotation-permutation
  problem via the IRA algorithm.}
  \label{fig:twostage}
\end{figure}

The workflow implements the Iterative Rotational Alignment (IRA)
\cite{gundeIRAShapeMatching2021} in two stages (Figure \ref{fig:twostage}):

\begin{description}
\item[{Pre-minimization}] Raw endpoints centered in \([25, 25, 25] \AA\) cell, aligned
using IRA with \(kmax=1.8 \AA^{-1}\) (configurable). Ensures consistent atom
ordering before geometry optimization.

\item[{Post-minimization}] After endpoint relaxation, IRA alignment reapplied to
correct atom mapping drift during minimization. Final RMSD (Root mean squared
deviation) between reactant and product logged for verification.
\end{description}

The \texttt{ira\_align} module employs the Iterative Rotational Alignment method
\cite{gundeIRAShapeMatching2021}, calculating optimal rotation and permutation to
minimize Hausdorff distance. For configurations \(\mathbf{X}\) with \(N\) atoms, the
permutation-invariant RMSD is:

\begin{equation}
d(\mathbf{X}, \mathbf{X}_{\text{ref}}) = \min_{\mathbf{Q} \in SO(3), \boldsymbol{\Pi} \in S_N} \sqrt{ \frac{1}{N} \left\| \mathbf{X} - \mathbf{Q} \mathbf{X}_{\text{ref}} \boldsymbol{\Pi} \right\|_F^2 }
\end{equation}

where \(\mathbf{Q}\) is the optimal rotation matrix and \(\boldsymbol{\Pi}\) the
permutation matrix. Automating this step ensures the path starts from the
shortest distance between endpoints in configuration space.
\subsection{Initial Path Generation}
\label{sec:org8d239df}

\begin{figure}[b]
  \centering
  \includegraphics[width=0.9\textwidth]{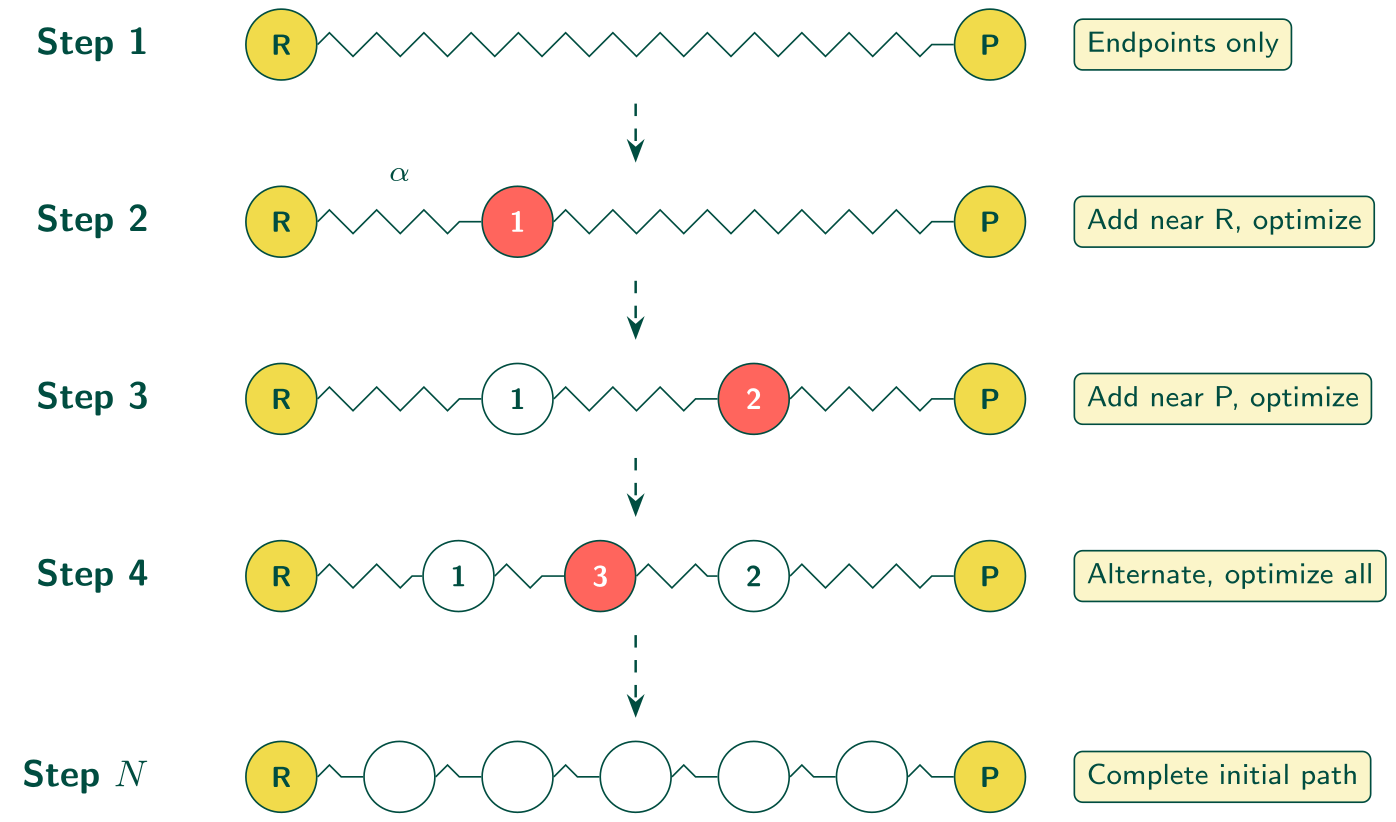}
  \caption{Sequential IDPP (SIDPP) path growth. Unlike standard IDPP which
  interpolates all images simultaneously, SIDPP adds one image at a time,
  alternating between reactant and product sides. After each addition, all
  intermediate images are re-optimized. The step size parameter $\alpha$ controls
  placement of each new image relative to the frontier. This sequential growth
  avoids local minima that trap simultaneous interpolation for complex reactions.}
  \label{fig:sidpp}
\end{figure}

With aligned endpoints, the workflow generates an initial collision-free path
using Sequential IDPP (SIDPP) \cite{schmerwitzImprovedInitializationOptimal2024}.
Unlike standard IDPP \cite{smidstrupImprovedInitialGuess2014} which interpolates
all images at once then optimizes, SIDPP grows the path sequentially (Figure
\ref{fig:sidpp}):

\begin{enumerate}
\item Start with only endpoints [R, P]
\item Add one image at a time (alternating reactant and product sides)
\item Optimize all intermediate images after each addition
\item Repeat until target image count reached
\end{enumerate}

Sequential growth with intermediate optimization avoids local minima that trap
standard IDPP for complex reactions. The step size parameter \(\alpha=0.33\)
controls placement of each new image relative to the frontier. Standard IDPP is
available as a fallback.
\subsection{Off-path CI-NEB (OCINEB) Optimization}
\label{sec:org46feb49}

The optimization uses Climbing Image NEB (CI-NEB)
\cite{henkelmanClimbingImageNudged2000} with energy-weighted springs
\cite{asgeirssonNudgedElasticBand2021} and minimum mode following (MMF) refinement
\cite{goswamiAdaptivePruningIncreased2025}.  For image \(i\) with positions
\(\mathbf{R}_i\), the NEB force is:

\begin{equation}
\mathbf{F}_i^{\text{NEB}} = -\nabla E(\mathbf{R}_i)|_{\perp} + \mathbf{F}_{i}^{\text{spring}}|_{\parallel}
\end{equation}

where \(\nabla E|_{\perp}\) is the potential force component perpendicular to the
path tangent and \(\mathbf{F}^{\text{spring}}|_{\parallel}\) is the spring force
projected along the tangent. The energy-weighted spring constant \(k_i =
k_{\min} + \Delta k \max(E_i - E_{\text{ref}}, 0) / (E_{\max} - E_{\text{ref}})\)
concentrates images near the barrier by stiffening springs between high-energy
images. The climbing image inverts the parallel force component at the
highest-energy image \(\mathbf{R}_{\text{CI}}\):

\begin{equation}
\mathbf{F}_{\text{CI}} = -\nabla E(\mathbf{R}_{\text{CI}}) + 2\left(\nabla E(\mathbf{R}_{\text{CI}}) \cdot \hat{\tau}\right)\hat{\tau}
\end{equation}

driving it uphill along the path tangent \(\hat{\tau}\) while relaxing
perpendicular to it. Once the climbing image converges, the workflow switches to
off-path climbing image NEB (OCI-NEB) \cite{goswamiEnhancedClimbingImage2026} with
MMF refinement along the lowest curvature mode.

Configuration parameters:

\begin{description}
\item[{Images}] 18 (including endpoints)
\item[{Maximum iterations}] 1000
\item[{Force convergence}] 0.0514221 eV/\AA{}$\backslash$ (10\(^{-3}\) Ha/Bohr) \cite{marksIncorporationInternalCoordinates2025}
\item[{SIDPP growth (\(\alpha\))}] 0.33
\item[{Climbing image activation}] 80\% convergence
\item[{Energy-weighted springs}] true (0.972--9.72 eV/\AA{}\(^2\))
\item[{Mode alignment with NEB tangent}] 0.8 (amount by which the rotated dimer can differ from)
\end{description}

All parameters are user-configurable, and Snakemake re-runs only those steps
whose inputs have changed. The complete set of parameters are part of the
accompanying repository and documented there.
\subsection{Applicability and Parameter Guidelines}
\label{sec:org8e09e70}

The workflow is optimized for gas-phase molecular systems with 5--50 atoms.
Table \ref{tbl:defParams} provides default parameters and tuning guidelines for different system
types.

\begin{table}[t]
\centering
\caption{Default parameters and tuning guidelines by system type}
\label{tbl:defParams}
\begin{tabular}{lccc}
\toprule
\textbf{Parameter} & \textbf{Small} & \textbf{Medium} & \textbf{Complex} \\
& \textbf{(5--20 atoms)} & \textbf{(20--50 atoms)} & \textbf{(50+ atoms)} \\
\midrule
NEB images & 12--16 & 18--24 & 24--30 \\
Force threshold (eV/\AA) & 0.051 & 0.051 & 0.026 \\
IRA kmax (\AA$^{-1}$) & 1.5--2.0 & 2.0--3.0 & 3.0--5.0 \\
Cell size (\AA) & 20--25 & 25--30 & 30--40 \\
\bottomrule
\end{tabular}
\end{table}

Parameter selection principles:

\begin{description}
\item[{Number of images}] Simple reactions (single bond break/form) need 12--16
images. Complex rearrangements with multiple bond changes require 24--30
images for adequate path resolution.

\item[{Force convergence}] Standard threshold (0.051 eV/\AA{}$\backslash$ = 10\(^{-3}\) Ha/Bohr)
suffices for most applications. High-precision kinetics studies should use
0.026 eV/\AA{}$\backslash$ (0.5\$\texttimes{}\(10\)\textsuperscript{-3}\$ Ha/Bohr).

\item[{IRA kmax}] Controls permutation matching strictness. Well-behaved small
molecules work with kmax=1.5--2.0 \AA{}\(^{-1}\). Difficult atom mapping
(symmetric molecules, multiple identical fragments) may require kmax=3.0--5.0
\AA{}\(^{-1}\). The 2D RMSD visualization uses fixed kmax=14 \AA{}\(^{-1}\) for
consistency across systems. The IRA procedure becomes infeasible for extended
systems in most cases.

\item[{Cell size}] Must be large enough to avoid periodic boundary artifacts. Default
{[}25, 25, 25] \AA{}$\backslash$ works for most medium-sized molecules. Surface reactions or
condensed-phase systems require modifications to handle periodic boundary
conditions correctly, which is part of eOn natively, and so the workflow extends to these systems without modification.
\end{description}
\section{Method Validation}
\label{sec:org8136bcf}

To demonstrate that the workflow produces correct results from raw inputs
without user intervention, we apply it to the HCN \(\to\) HNC isomerization, a
3-atom proton transfer with a single transition state. CCSD(T)/cc-pCV5Z
calculations \cite{vanmourikInitioGlobalPotential2001} place the barrier at 2.09
eV and the HCN--HNC energy difference at 0.65 eV.

\begin{figure}[t]
  \centering
  \includegraphics[width=\textwidth]{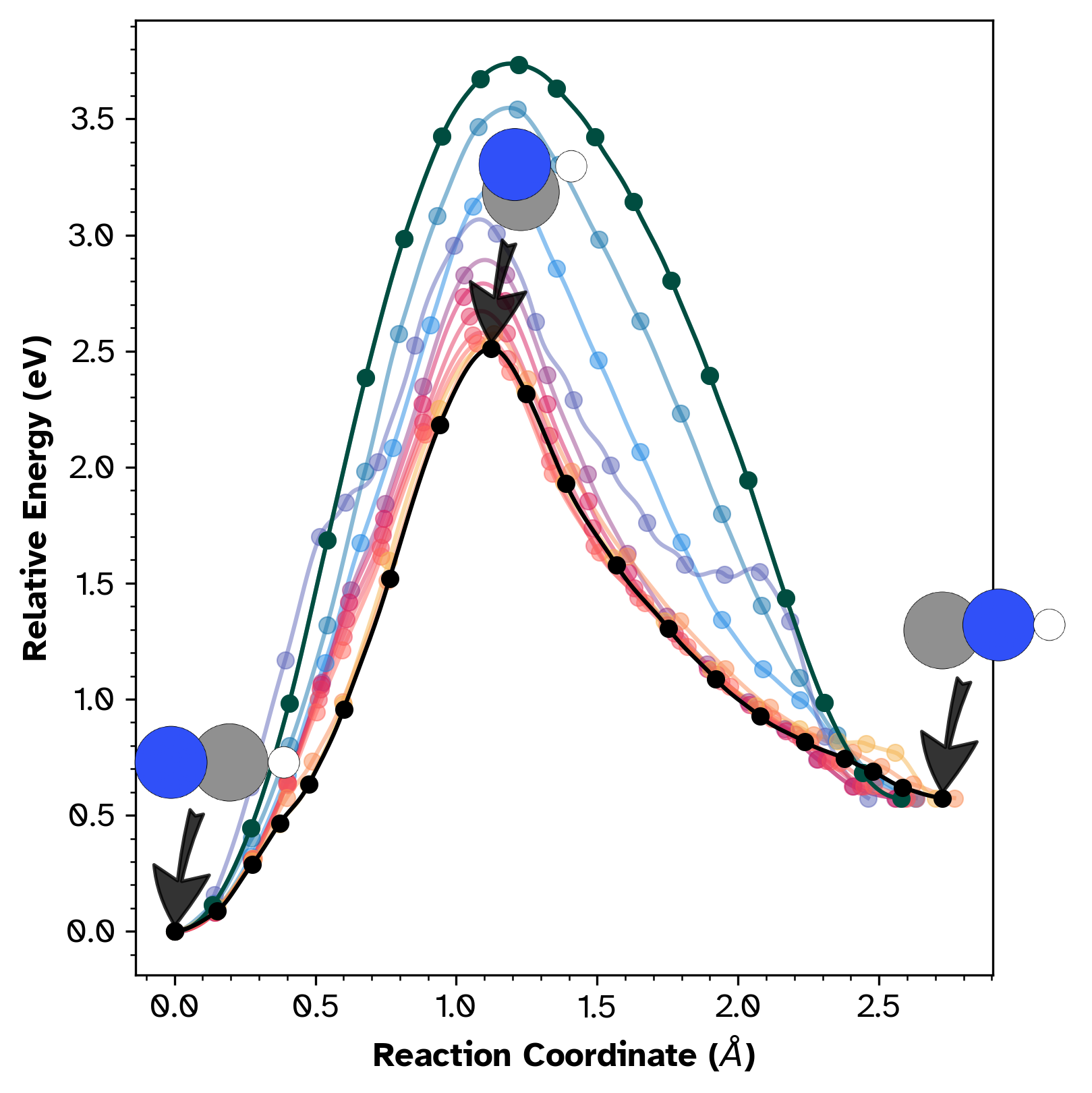}
  \caption{One-dimensional energy profile for the HCN $\to$ HNC
  isomerization. The single barrier of 2.46~eV separates the HCN reactant
  from the HNC product (0.57~eV above HCN). Colored traces show the
  optimization history from initial SIDPP guess (outer traces) to the
  converged path (black). Inset structures show the reactant, saddle point,
  and product geometries.}
  \label{fig:profile}
\end{figure}

\begin{figure}[b]
  \centering
  \includegraphics[width=\textwidth]{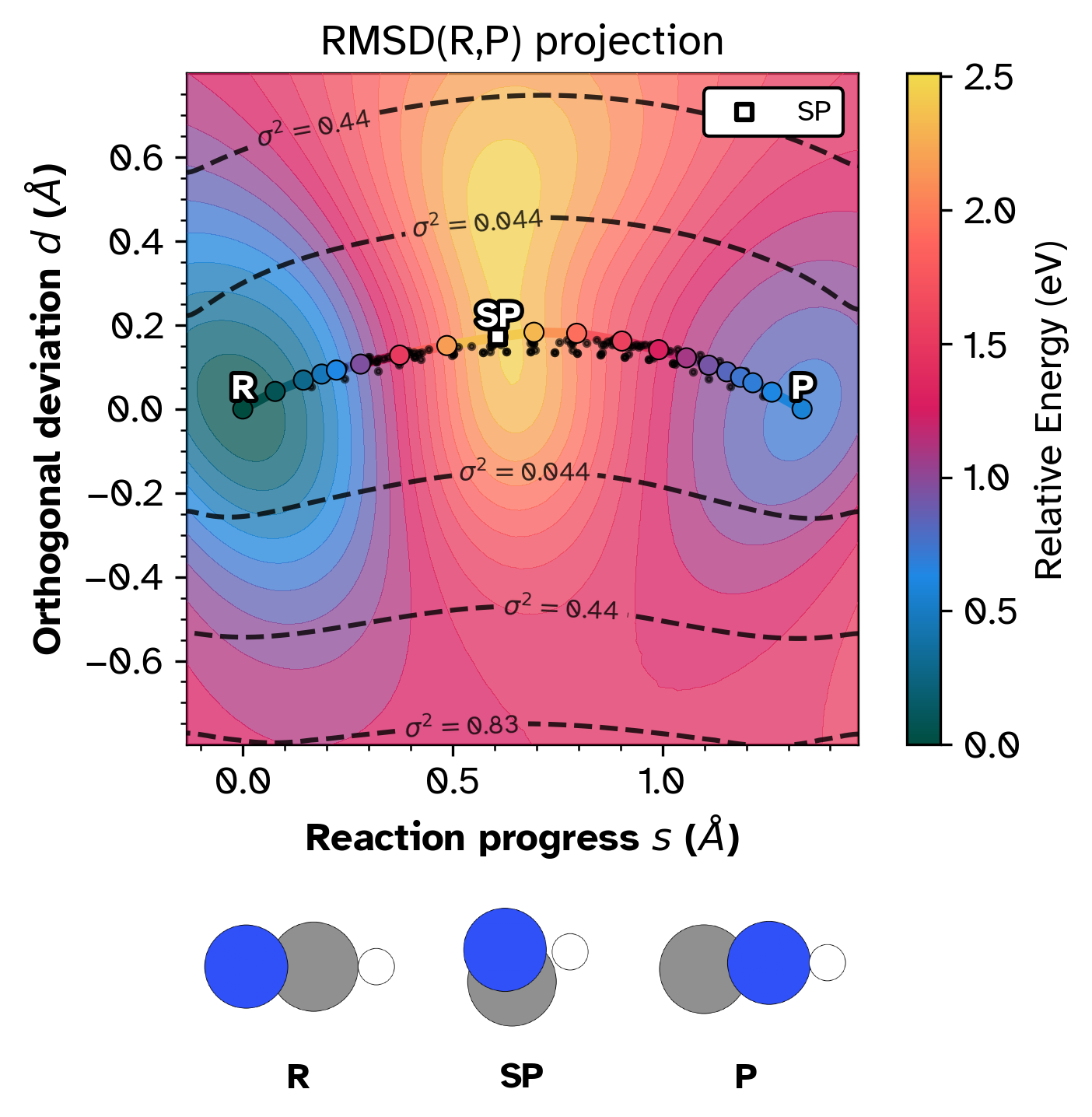}
  \caption{Two-dimensional RMSD landscape for the HCN $\to$ HNC isomerization.
  Axes represent permutation-invariant RMSD from reactant (reaction progress
  $s$) and orthogonal deviation ($d$). The color scale indicates interpolated
  energy. The reactant (R), saddle point (SP), and product (P) basins are
  clearly resolved.}
  \label{fig:landscape}
\end{figure}

Starting from endpoint structure files alone, the workflow completes
minimization, IRA alignment, SIDPP path generation, and hybrid CI-NEB-MMF
optimization. The resulting energy profile (Figure \ref{fig:profile}) recovers the
expected single-barrier topology with a barrier of 2.46 eV and a product energy
0.57 eV above the reactant. The 2D RMSD landscape (Figure \ref{fig:landscape})
resolves the reactant, saddle point, and product basins.

\begin{figure}[b]
  \centering
  \includegraphics[width=\textwidth]{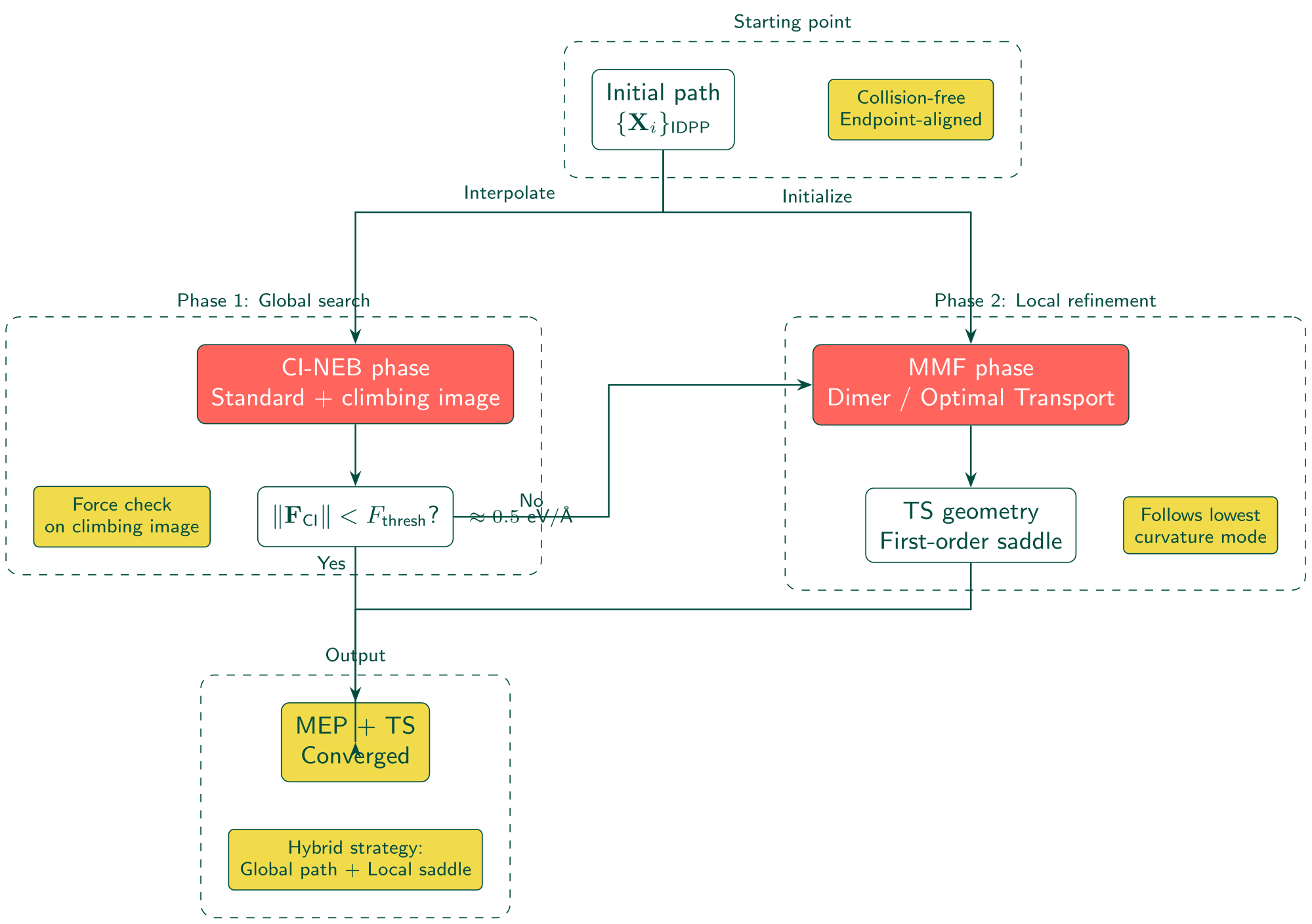}
  \caption{Hybrid CI-NEB-MMF optimization strategy. The CI-NEB phase performs
  global path optimization with a climbing image. When the climbing image force
  falls below threshold ($\approx 0.5$ eV/\AA), the method switches to MMF for
  local saddle refinement via the lowest curvature mode.}
  \label{fig:hybrid}
\end{figure}

The repository includes four additional systems: the alanine dipeptide C7\textsubscript{eq}
\(\to\) C5 conformational transition, Diels-Alder cycloaddition of ethylene and
butadiene, the SN2 identity reaction F\textsuperscript{-} + CH\textsubscript{3F}, and vinyl alcohol to
acetaldehyde tautomerization. Each requires only an endpoint pair and a YAML
configuration file; the workflow handles alignment, path generation, and
optimization identically. All five systems converge without manual intervention,
and adding a new system amounts to providing two endpoint structures and editing
a single configuration file.
\section{Reproduction}
\label{sec:org41aa09b}

Complete reproduction proceeds as follows:

\begin{lstlisting}[language=bash,numbers=none]
git clone https://github.com/HaoZeke/eon_orchestrator
cd eon_orchestrator
pixi shell -e eon
snakemake --configfile examples/hcn_isom/config.yaml -c4
\end{lstlisting}

This generates plots and intermediates in the results folder. To reproduce the
exact figures:

\begin{lstlisting}[language=bash,numbers=none]
# Figures 1-4: TikZ schematics (pre-built)
# See paper/imgs/tikz/ in the manuscript repository

# Figures 5-6: Energy profile and 2D landscape
# Generated automatically in results/plots/hcn_isom/
ls results/plots/hcn_isom/
# 1D_path.png (Figure 5) and 2D_rmsd.png (Figure 6)
\end{lstlisting}

Documentation accompanies the repository with tutorial guidelines.
\section{Visualization Framework}
\label{sec:org52d11cc}

The workflow produces two complementary visualizations.
\subsection{One-Dimensional Energy Profiles}
\label{sec:org1ac1777}

Energy is plotted against a scalar reaction coordinate. Three variants are
available:

\begin{description}
\item[{Image index}] Simplest, plots \(E_i\) vs. \(i \in [0, N]\). Loses all geometric
information about inter-image spacing.

\item[{Cumulative path length}] \(s_i = \sum_{j=1}^{i} \| \mathbf{X}_j -
  \mathbf{X}_{j-1} \|_2\). Accounts for non-uniform image distribution but
remains path-dependent.

\item[{RMSD from reactant}] \(r_i = d(\mathbf{X}_i, \mathbf{X}_\text{react})\). Uses
permutation-invariant RMSD; different methods can be compared on the same
axis.
\end{description}

All profiles use the tangential forces \(F_{\parallel, i} = -\mathbf{F}_i \cdot
\hat{\tau}_i\) to constrain the energy derivative during interpolation, producing
physically consistent barriers \cite{henkelmanClimbingImageNudged2000}.
\subsection{Two-Dimensional RMSD Landscapes}
\label{sec:orgc4e861c}

One-dimensional projections collapse the \$3N\$-dimensional configuration space
onto a single axis, obscuring path tortuosity and basin topology. The 2D
framework projects onto intrinsic coordinates defined by the endpoints
themselves \cite{goswamiTwodimensionalRMSDProjections2026}.

Using the permutation-invariant RMSD from Eq. (1), we compute distances from
each configuration \(\mathbf{X}\) to both reactant (\(\mathbf{R}\)) and product
(\(\mathbf{P}\)). The raw coordinates \((r, p) = (d(\mathbf{X}, \mathbf{R}),
d(\mathbf{X}, \mathbf{P}))\) form a triangular region in the RMSD plane. A rigid
rotation decomposes these into:

\begin{align} s &= \text{reaction progress (along path)} \\ d &=
\text{orthogonal deviation (perpendicular to path)} \end{align}

This decomposition separates forward progress from lateral excursions, analogous
to path collective variables \cite{branduardiFreeEnergySpace2007} but without
smoothing parameters or training requirements.

The 2D landscape exposes path tortuosity, intermediate basins, and stagnation
regions that 1D profiles collapse into featureless shoulders.  The projection
operates directly on Cartesian coordinates without training data; the \(\approx
10^3\) geometries from a single NEB run suffice.  Visualization code is provided
in the \texttt{chemparseplot} and \texttt{rgpycrumbs} Python packages.
\section{Limitations}
\label{sec:org6e84099}

\begin{description}
\item[{ML potential accuracy}] PET-MAD-XS provides point estimates without
uncertainty quantification. Barrier heights should be validated with DFT
single-point calculations for quantitative kinetics. Systems outside the
training domain (transition metals
\cite{schreinerTransition1xDatasetBuilding2022}, long range effect systems
\cite{locheFastFlexibleLongrange2025}) may require model finetuning.

\item[{Permutation sensitivity}] Two-stage IRA alignment ensures consistent endpoint
mapping but does not guarantee correct permutations along the entire path.
Reactions with many equivalent atoms or soft rotational modes warrant visual
inspection of the initial path \cite{goswamiEnhancedClimbingImage2026}.
\end{description}

These limitations reflect the current state of ML potentials and
permutation-invariant alignment, not the workflow itself.
\section{Conclusions}
\label{sec:org094249a}

The workflow automates NEB calculations from raw endpoint structures through
publication-quality visualizations, with every step declared as a node in a
Snakemake DAG. The explicit dependency graph removes the manual setup that
causes most NEB failures in practice: mismatched atom orderings, un-minimized
endpoints, and poorly initialized paths. All dependencies resolve from
conda-forge via pixi, so reproduction requires four commands. The workflow
accepts any eOn compatible potential including the many metatomic interfaces
\cite{bigiMetatensorMetatomicFoundational2026}  and runs unchanged on a laptop or
HPC cluster.
\section{CRediT Author Statement}
\label{sec:orgec7ad5d}

Rohit Goswami: Conceptualization, Methodology, Software, Visualization,
Writing - Original draft preparation, Writing - Reviewing and Editing
\section{Acknowledgments}
\label{sec:org3565382}

R.G. acknowledges support from his family, particularly Ruhila Goswami, Prof. D.
Goswami, S. Goswami, Dr. A. Goswami, Dr. M. Sallermann and co-authors on the
metatensor ecosystem publication. R.G.  also acknowledges support from Lab-COSMO
at EPFL, in particular Prof. Ceriotti and Dr. Fraux. R.G. would like to
specially thank his pets, plants, birds, garden cats, turtles, and other
constant supportive help. Dedicated to the memory of Snoopy.

This research did not receive any specific grant from funding agencies in the
public, commercial, or not-for-profit sectors.
\section{Data Availability}
\label{sec:org2193d28}

The workflow source code is available at
\url{https://github.com/HaoZeke/eon\_orchestrator} under the MIT license. A complete
data archive containing all computed results, the ML potential model,
per-iteration NEB data, and publication-quality figures is deposited on the
Materials Cloud

Archive. Extracting the archive at the repository root recreates the exact
computational state used in this work. See the included \texttt{README.md} for
the data layout and file format descriptions. This resource is expected to
facilitate rapid exploration of energy surfaces with scalable reproducible
workflows. Part of the concepts described here are part of the \href{https://atomistic-cookbook.org/examples/eon-pet-neb/eon-pet-neb.html}{atomistic
cookbook} as well.
\section{Declarations}
\label{sec:orgbcf699c}

\subsection{Ethics statements}
\label{sec:org62e4c39}

Not applicable.
\subsection{Declaration of interests}
\label{sec:org6bcb368}

The authors declare that they have no known competing financial interests or
personal relationships that could have appeared to influence the work reported
in this paper.

\bibliography{roneb_orch2026}
\end{document}

%% file: authors_arxiv.tex
\author{ \href{https://orcid.org/0000-0002-2393-8056}{\includegraphics[scale=0.06]{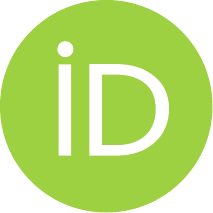}\hspace{1mm}Rohit Goswami} \\
  Institute IMX and Lab-COSMO \\
  \'Ecole polytechnique f\'ed\'erale de Lausanne (EPFL) \\
  Station 12, CH-1015 Lausanne \\
  Switzerland\\[1ex]
  \texttt{rohit.goswami@epfl.ch}
}